\documentstyle[twocolumn,aps]{revtex}
\input{epsf.sty}

\begin{document} 
\draft
\title{Coulomb Blockade and Cotunneling in Single Electron Circuits 
with On-chip  Resistors: Towards the Implementation of R-pump}
\author{ A.~B.~Zorin, S.~V.~Lotkhov, H.~Zangerle and  J.~Niemeyer\\}
\address{Physikalisch-Technische Bundesanstalt, Bundesallee 100, 
D-38116 Braunschweig, Germany\\}
\maketitle

\begin{abstract}
We report on the investigation of Al single electron structures
equipped with miniature (8~$\mu$m long) on-chip Cr
resistors of $R > R_k = h/e^2 \approx 26 \,$k$\Omega$.
From the measurement of the Coulomb blockade in single-junction 
structures we evaluated the self-capacitance of
our resistors per unit length, $c \approx 62$~aF/$\mu$m.   
We demonstrate that the cotunneling current in the 
transistor samples in the Coulomb blockade regime obeys
the power law, $I \propto V^{3+(R/R_k)}$,  
predicted by Odintsov, Bubanja and Sch\"on for a 
transistor having pure ohmic-resistance leads. 
The concept of the three-junction single electron pump
with on-chip resistors (R-pump) is developed. We demonstrate
that the implementation of the R-pump with a relative accuracy 
of the electron transfer of $10^{-8}$ is quite feasible 
with the technology available.
\end{abstract}

\pacs{PACS numbers: 73.40.Gk, 73.23.Hk}

\section{Introduction}

After the pioneering work by Averin and Likharev \cite{AvLik},
predicting the single electron tunneling (SET) effects in 
current-biased junctions of small intrinsic capacitance,
the series of experimental papers on metallic tunnel junction circuits 
with on-chip resistors was published  
(see, for example, Refs. \cite{Dels,Clel,KuzPash,Joyez,Zheng}).
The miniature resistors make it possible to realize 
a high environmental impedance 
$ |Z(\omega)| \gg R_k \equiv h/e^2 \approx 25.8~$k$\Omega$
in rather wide frequency range.
Such an isolation of the tunnel junctions
from a power source generally results in an enhancement of the 
Coulomb blockade in the $I-V$ curve of a SET device. \cite{Ing}
In particular, a high-ohmic compact resistor $R$ connected
in series with a single tunnel junction ensures that
the pronounced Coulomb blockade is observed 
(see, for example, Ref. \cite{Zheng}).
Moreover, a sufficiently large $R$   
allows the coherent SET \cite{AvLik} and Bloch \cite{LikZor} 
oscillations to be observed \cite{KuzPash,KuzHav} in these 
current-biased circuits.  

There is still another important advantage of the SET 
circuits with high-ohmic resistors, which up to date has 
not been tested in practice. Namely, in the Coulomb blockade 
state, these circuits can have an extremely low cotunneling 
current (i.e. the current component associated with 
coherent tunneling of an electron simultaneously over several 
junctions \cite{AvOd}). 
In the process of cotunneling, the electromagnetic
environment with large dissipation (a resistor of 
magnitude $R > R_k$) can absorb a considerable part of the electron
energy $eV$, which then is distributed over the numerous internal 
degrees of freedom of this dissipative environment. As a 
result, the cotunneling probability is drastically reduced.

The theory of cotunneling in $N$-junction arrays with resistors 
has been developed by Odintsov, $et$ $al.$ \cite{OBS} 
for $N=2$ 
and generalized by Golubev and Zaikin \cite{GolZai} for $N \geq 2$.
They showed that at $T=0$ the cotunneling contribution to 
the $I-V$ curve in the Coulomb blockade regime obeys the power law
$$ I \propto V^{\eta}, \qquad \eta = 2(N+z)-1, \eqno (1) $$
with the dimensionless parameter $z = R/R_k$. In this paper we report 
on the first experimental
observation of this effect in the normal Al SET structures with 
compact Cr resistors and discuss its possible impact on
the performance characteristics of SET devices and
the three-junction SET pump in particular. 

\section{Experiment}

\subsection{Fabrication technique and sample layout}

The metallic SET structures were deposited on the
thermally oxidized surface (SiO$_2$ layer of thickness 
of ca. 300 nm) of a Si wafer (about 300 $\mu$m thick). 
The tunnel junctions and microstrip resistors 
were fabricated $in$ $situ$
by the shadow evaporation technique \cite{NiemDol} through the
trilayer mask patterned by e-beam lithography and reactive-ion 
etching.
At three different angles, we then deposited three metal
layers: Cr (8-10 nm thick), then Al (30 nm), and, after oxidation, 
again    
Al (35 nm). The tunnel junctions had nominal 
dimensions of 80~nm by 80~nm,
resulting in the junction capacitance $C_j \approx 0.15-0.2$~fF
(or the junction charging 
energy $E_c = e^2/2C_j \approx 0.4-0.5$~meV) 
and tunnel resistance $R_j \approx 60-250$~k$\Omega$. 
An example of the resulting  structure, a SET transistor
with resistors (we will call it R-SET) 
is depicted in Fig.~\ref{SEM}.

The Cr microstrips (shown only partially in the SEM 
photo, Fig.~\ref{SEM}a) had 
the nominal lateral dimensions of $w=80$~nm by $ \ell =8$~$\mu$m 
and a resistance per unit length $r$ of up to 
10~k$\Omega$/$\mu$m (or $R_{\Box} = r w$ up to 800~$\Omega$ per 
square).
The total dc resistance $R \approx 30-80$~k$\Omega$
of Cr strips normally did not exceed $R_j$. 
(This ensured establishing of the charge
equilibrium before a tunneling event occurs, i.e. the
condition assumed in theory \cite{OBS}.)
On the other hand, these resistance magnitudes 
corresponded to substantial values of 
parameter $z \approx 1-3$. 
These values should ensure the strong suppression of
cotunneling, which, according to
Eq.(1), is equivalent to the action of one to three additional tunnel
junctions attached to the original structure.  

\subsection{Characterization of resistors}

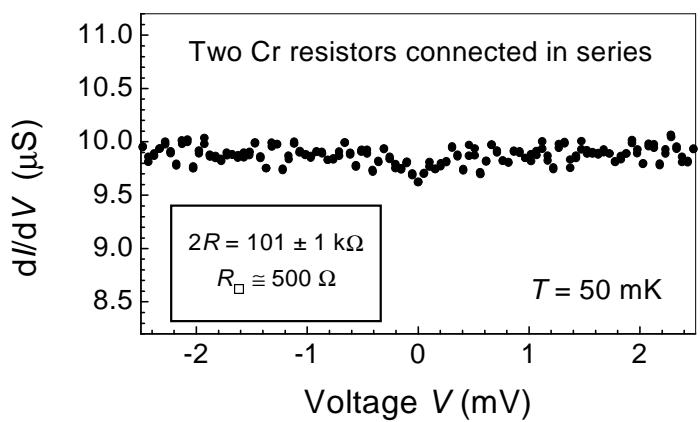
\begin{figure}
\caption{The derivative of the current-voltage characteristic 
of two Cr resistors connected in series. The curve shows
insignificant traces of the Coulomb blockade
which may be due to the granular nature of the film.} 
\label{dIdV}
\end{figure}

The resistors exhibited practically linear $I-V$ 
characteristics at temperatures down to the lowest 
temperature of our measurements $T=15$ mK. An example 
of the derivative plot $dI/dV-V$ is presented in 
Fig.~\ref{dIdV}. The strictly linear behavior of
our resistors (at least those having $R \leq 50$~k$\Omega$ 
or $R_{\Box} \leq 500~\Omega$ per square) compares 
favorably with that of the long arrays of tunnel 
junctions \cite{Dels}, 
which can hardly model the real metallic resistance 
behavior. Due to the inherent Coulomb 
blockade, these arrays usually
exhibit considerable non-linearity at small voltages,
that limits their application.

The microstrip resistor generally presents 
a lossy transmission line with an inductance 
per unit length,  ${\cal L} \sim \mu_0 \approx 1.26$ pH/$\mu$m
which reactance at characteristic frequencies of 10-100~GHz is
negligibly small, $< 1~\Omega / \mu m \ll r$.
The distributed capacitance of this $RC$-line was found 
experimentally from 
the normal-state $I$-$V$ characteristics of the single-junction 
structures with such resistors. 
Our idea was based on the fact that at $T = 0$ 
the second derivative of
the $I$-$V$ curve should 
mimic the function $P(E)$ determined by 
impedance $Z(\omega)$ (see, for example, 
the review paper \cite{IngNaz}), 
$$ {d^2I \over dV^2} = {e \over R_j} P(eV). \eqno (2) $$
Function $P(E)$ plays an important role in 
the "environment" theory of tunneling  where it 
describes the probability of the transfer of energy $E$ 
to the external circuit in a tunneling event. 

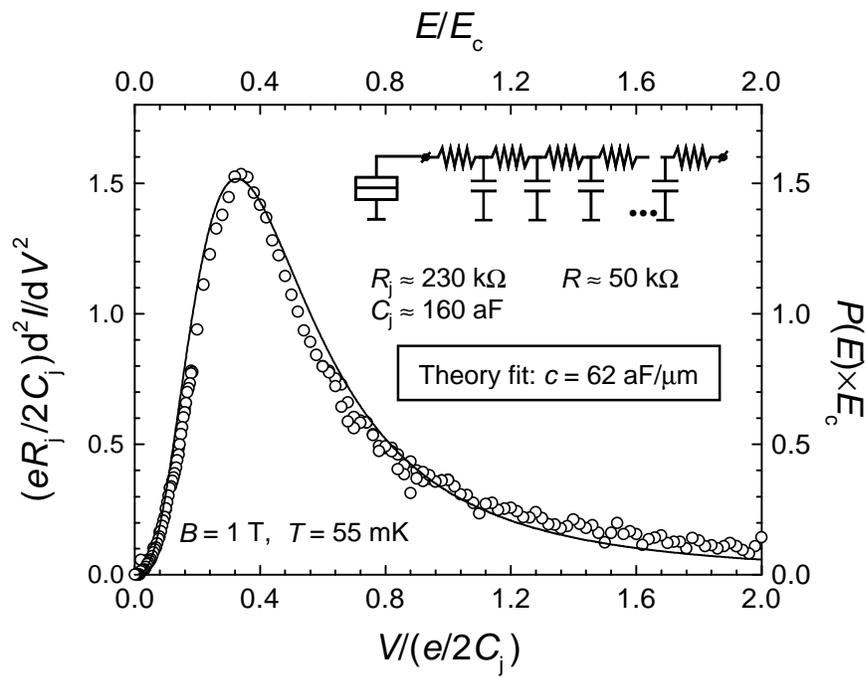
\begin{figure}
\caption{The second derivative plot (symbols) of 
the current-voltage characteristic of an Al single junction
equipped with four on-chip Cr resistors (similar to those fragmentarily 
shown in Fig. 1). The equivalent impedance of the junction environment 
is determined by the impedance of a single strip whose equivalent 
electric circuit is shown in the inset. Function $P(E)$ (solid line) 
was computed for an $RC$-line impedance for the given
(measured) $r = R/ \ell$ and fitted $c$. } 
\label{fit}
\end{figure}

First we measured the $dI/dV$-$V$ curves and 
then derived the $d^2I/dV^2$-$V$ curves and
averaged them over several runs.
Then we fitted the obtained curves with those computed from 
theory \cite{IngNaz} for the particular case of a 
$RC$-line impedance.
The line resistance $r$ was taken from the dc measurements 
while the
line capacitance per unit length $c$ was the fitting parameter.
The resulting fit is shown in Fig.~\ref{fit}.
It gave the value of specific capacitance 
$c \approx 62$~aF/$\mu$m \cite{Kuz}, or total capacitance 
$C_{0} = c \, \ell \approx 500$~aF of the whole line.
A similar estimation, $c \approx 60$ aF/$\mu$m, was 
made by Kuzmin $et$ $al.$ \cite{Kuz} from the 
shape of the supercurrent bump in the $I-V$ 
characteristic of a small Josephson junction
attached to 0.1~$\mu$m-wide and 28 $\mu$m-long 
Cr resistors. 

\subsection{Measurements of cotunneling current}

For evaluating the effect of cotunneling in Al-Cr
circuits, we performed measurements of the $I-V$ curves 
of R-SETs. Qualitatively similar to $I-V$ curves of
SET transistors without resistors, the $I-V$ curves of
our structures were featured by 
a sharp Coulomb blockade corner and a steep decrease 
of $I$ inside the blockade region (typical of SET arrays 
with a large number of junctions $N > 2$). This pointed 
to a strong suppression of the cotunneling current due 
to the attached resistors.

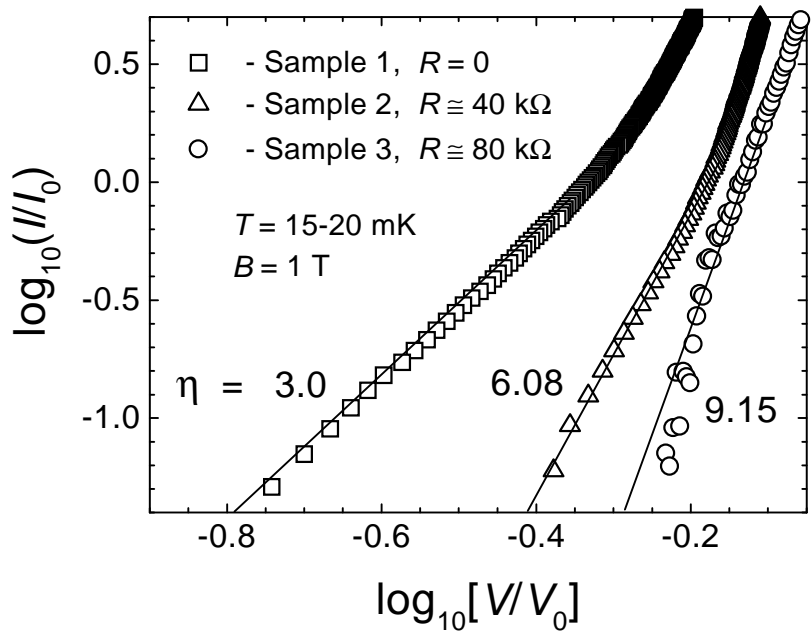
\begin{figure}
\caption{
$I-V$ characteristics (points) of three Al SET transistors in the 
Coulomb blockade regime in the normal state. 
The bare sample 1 was not equipped with resistors. 
Either of samples 2 and 3 (see their micrograph in Fig. 1) has 
four similar Cr microstrip resistors with a magnitude 
of about 40~k$\Omega$ and 80~k$\Omega$, respectively.
The gate voltages were adjusted
to maximize the Coulomb blockade. 
The voltage ($V_0$) and current ($I_0$) units are given 
in Table I. 
The solid straight lines
show the theoretical dependence given by Eq.(1) for the 
values of $\eta = R/R_k$ taken from dc measurements of $R$. }
\label{cotunn}
\end{figure}

To obtain higher accuracy of measurement inside the Coulomb
blockade region, the lock-in technique was applied. 
The log-log plot of the $I-V$ curves of three different 
samples is presented in Fig.~\ref{cotunn}.
The electric parameters of these samples are compiled
in Table I. The reference sample 1 (without resistors) shows 
a cubic decrease of current, $\eta=3$, which is typical of the 
low-temperature behavior of ordinary SET transistors \cite{Geer}.
The experimental data for R-SET samples 2 and 3 with 
resistors of 40~k$\Omega$ and 80~k$\Omega$, respectively,
measured in the range 0.2~mV $<V<$ 0.4~mV demonstrate a 
much steeper decay, $\eta =2z + 3 \approx$~6-9. 

\section{Suppression of cotunneling by on-chip resistors}

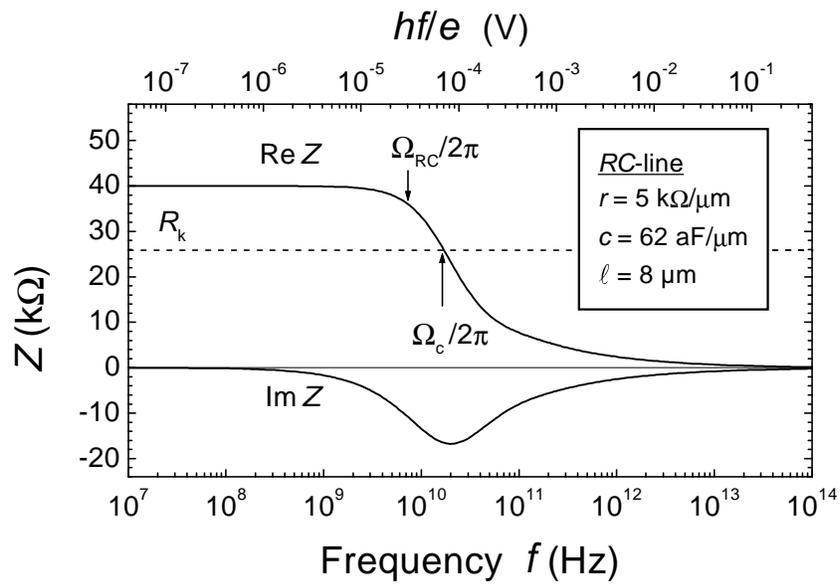
\begin{figure}
\caption{The electromagnetic impedance of a $RC$-line
modeling the typical Cr resistor (similar to that
of transistor sample 2) versus frequency.
The upper scale is the conversion of frequency into the voltage 
units. 
The characteristic frequencies $\Omega_{RC}$ and $\Omega_c$ (see 
text) are shown by arrows. The quantum resistance level is
shown by a dashed line.} 
\label{Zw}
\end{figure}

We found that the behavior of R-SETs is well
described by Eq.(1) in 
the range of the bias voltage up to $V \approx
200-400~\mu$V. 
This fact is still surprising because Eq.(1)
was derived for the pure ohmic impedance, which for our samples 
is realized for substantially lower voltages,
$V < V_{RC} = \hbar\Omega_{RC}/e \approx 15-30 \; \mu$V, where 
$$\Omega_{RC}= {1 \over RC_{0}} 
\approx (2.4-4.8) \times 10^{10} {\rm~s}^{-1}, \eqno (3)$$
the roll-off frequency of the transmission line (see Fig.~\ref{Zw}).
Moreover, the range of applicability of Eq.(1) was found to be 
extended even somewhat
beyond the voltages $V_c = \hbar\Omega_c/e$, where 
$$\Omega_c = {8r \over c R_k^2} \approx (1-2) 
\times 10^{11} {\rm~s}^{-1} , \eqno(4) $$
the frequency at which the real part of the line impedance
drops down to the level Re$Z(\omega) = R_k$ 
(see Fig.~\ref{Zw}). 
In our case, $V_c \approx 70~\mu$V and $140~\mu$V  for
samples 2 and 3, respectively.

The data show, that the effect of the resistive microstrips is 
very close to that of pure resistors in substantially wider
than expected range of voltage. This disparity 
between the encouraging 
experimental results and rather conservative predictions of 
theory \cite{OBS} is most likely because of the 
theoretical model which does not take into account a
non-ohmic dispersion of real resistors. That is why a 
more general approach (which deals with the $RC$-line environment) 
is urgently needed. Such theory should
give the quantitative  
answer to the question what the optimum length $\ell$ of 
a resistor is which causes an effective suppression of cotunneling 
in the whole range of the Coulomb blockade.
So far we accept the empirical criterion $\ell \sim 3C_j/c$.

Another important experimental issue is the measurement of  
very low cotunneling current. The accuracy of both dc and 
lock-in measurements is severely limited by the noise of preamplifier.
However, there is more efficient way to evaluate very small
$I$: it is to count individual electrons carrying this current.   
The experiment of this kind was recently carried out by 
Lotkhov $et$ $al.$ \cite{Trap} with
a four-junction chain of tunnel junctions attached on one 
side to a voltage source through a 50~k$\Omega$ resistor 
(identical to those described in this paper) and, on the other side, to a 
so-called memory island. The change of 
the number of electrons on the memory island was reliably
measured by a SET electrometer capacitively  coupled to
that island. Due to the presence of the resistor,
this so-called R-trap demonstrated an amazing 
capability of holding the electric charge (or, in other terms, of not 
leaking) over hours. This characteristic
is more than three orders of magnitude better that that of
the ordinary four-junction chains without resistors. \cite{4trap}
In fact, it is 
comparable to that of the devices consisting of $N=$ 6-7 
junctions \cite{Kell,Dress} in which the cotunneling is
drastically suppressed because of a large number of junctions in
the chains.
Thus, the experiment \cite{Trap} complements the present
study of cotunneling current suppression from the side of very
low currents.

\section{Towards the R-pump}

The results of our experiments encourage us to further develop 
the family of fewer-junction SET circuits with on-chip resistors. 
These circuits can have an extremely small cotunneling component 
while the sequential tunneling component of SET
can still be efficiently controlled by gates. 
The most attractive devices are the SET turnstile and pump
with local resistors, or R-turnstile and R-pump, respectively.
They are capable of carrying an accurate 
dc current $I= ef$, where $f$ is the frequency of the rf drive applied 
to the gate(s). These devices were first referred to 
Odintsov $et$ $al.$ \cite{OBS}; here we develop a concept for
their operation and make a preliminary 
evaluation of the characteristics expected.

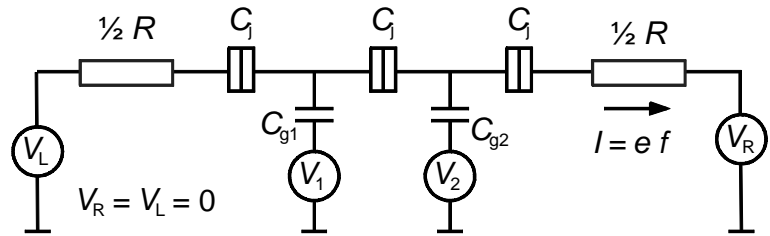
\begin{figure}
\caption{The equivalent electric circuit diagram of a three-junction
R-pump. 
}
\label{Rpump-eqv}
\end{figure}

As an example, Fig.~\ref{Rpump-eqv} illustrates the basic 
electric circuit of a three-junction R-pump. It comprises two islands
supplied with the gates of capacitances $C_{g1}, C_{g2} \ll C_j$.    
These gates are driven by two harmonic signals,
$$ V_{1,2} = V_{01,02} + A \cos(2\pi f t \pm {\theta \over 2}) \eqno(5)$$
of the frequency $f \ll  (R_jC_j)^{-1}$ and a relative phase 
shift $\theta$. Voltages $V_{01,02}$ adjust the working 
point ensuring the pumping regime.
Varying $\theta$ in the range $[90^{\rm{o}}, 180^{\rm{o}}]$ one 
can compensate an unavoidable cross-talk between the gates in such
a way that polarization charges on the two islands alternate in time
as quadrature components. This phasing ensures the optimum pumping
cycle (see, for example, the review by Esteve \cite{Est}), whose
parameters, as will be shown below,  remaining intact in our case. 
The key elements of the R-pump are two resistors, $R/2$ each, 
$$R_k < R < R_j, \eqno(6) $$
and this condition ensures establishing of the charge equilibrium 
in the circuit at the rate $\tau^{-1} \sim (RC_j)^{-1} \gg f$. 
This means that the resistors carry current only during a short
time $\sim \tau$ after a tunneling event. 
Then the voltage drop across the resistors is zero and
the chain of junctions is at zero bias, $V_L=V_R=0$,
similarly to the case of $R=0$.

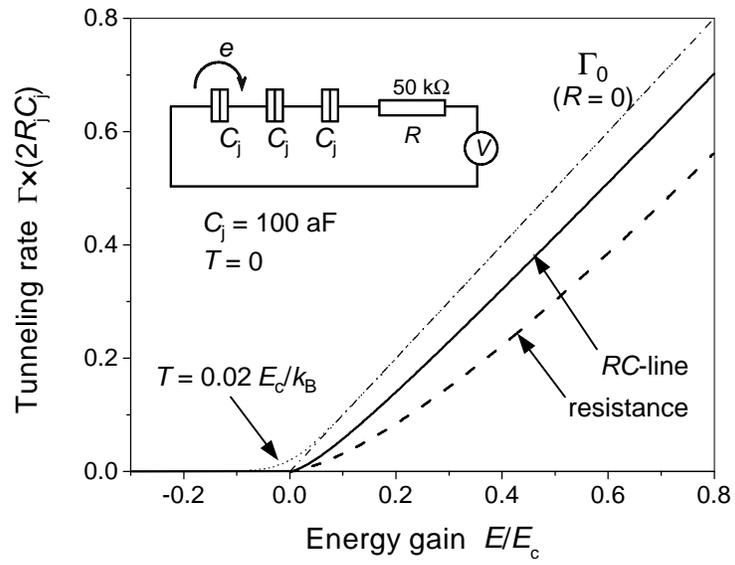
\begin{figure}
\caption{
The rate of electron tunneling through a particular
junction of R-pump (modeled by the simplified network shown
in the inset) versus the energy gain associated with this
tunneling. The dashed line relates to a pure ohmic 
resistance of 50~k$\Omega$, while the 
solid line gives the data for the $RC$-line of similar dc
resistance. The dash-dotted
and dotted lines describe the ordinary case of the zero
resistance $R$ for the zero and finite temperature, respectively.
The curves for $R \neq 0$ should be changed with temperature
in a qualitatively similar way. 
}
\label{Rpump-rate}
\end{figure}

The most important issue in the pump dynamics is the
dependence of the electron tunneling rate $\Gamma$ on 
the energy $E$ gained in this event. In our case, $\Gamma$ 
is expressed as a convolution integral \cite{IngNaz}
$$ \Gamma (E) = \int_{-\infty}^{\infty} 
\Gamma_0 (E') \, P_3(E-E')\, dE', \eqno (7)$$
where
$$\Gamma_0 (E) = {1 \over {e^2R_j}} 
{E \over 1-\exp(-E/k_BT)}	\eqno (8)$$
is the usual rate for $R=0$. 
Similar to function $P(E)$ of a single-junction 
circuit with resistor, the "environment function" $P_3(E)$ 
describes the absorption of energy in the three-junction 
circuit with resistor. In contrast to  $P(E)$ 
(see Fig.~\ref{fit}), $P_3(E)$ is generally a "more peaked" 
function with the peak position closer to the origin $E = 0$.
This is because two additional junctions connected in series 
with resistor $R$ (see the inset of Fig.~\ref{Rpump-rate})
considerably (by the factor of $\kappa^2_1 = (1/3)^2$) 
reduce the dissipative part of the impedance 
seen by the given junction. 
For not very large $R$, 
say, $\sim (2-4) R_k \approx 50-100$~k$\Omega$, 
this results in a dependence $\Gamma (E)$ which is not very much
different from $\Gamma_0 (E)$. This is 
illustrated in Fig.~\ref{Rpump-rate} 
for the ohmic resistor as well as for the $RC$-line.
One can see that at $T=0$ all three curves have the same
onset threshold $E=0$. 

From this consideration we can draw two important conclusions.
First, the stability diagram of the R-pump in the plane of 
gate voltages {$V_{1}, V_{2}$} is the same as that of 
the ordinary pump \cite{Est}, i.e. it consists of hexagonal 
domains with the triple-points.
The coordinates of these nodes are identical for the two devices,
so there is no need to revise the pumping cycle parameters
$V_{01}, V_{02}, A$ and $\theta$. 
Secondly, the curves computed for $\Gamma$ in the range of 
characteristic values of energy, $0 < E < \tilde{E} \sim  0.2 E_c$  
(dependent on the cycle parameters) fall below the 
dependence $\Gamma_0 = E/(e^2R_j)$ given by Eq.(8) at $T=0$ 
and can be roughly approximated as $\Gamma \approx E/(e^2R')$
with $R' \sim 2R_j$ and $3R_j$ for the $RC$-line and
the pure resistor, respectively. This means
that the resistors somewhat hamper (slow down) the 
sequential electron tunneling in the R-pump, and a reduction of 
frequency $f$ might be required to ensure reliable operation. 

To minimize the errors $\delta I_{\rm{mis}}$
associated with missing the tunneling events, 
the operation frequency $f$ should be rather low. \cite{Est}
For the ordinary pump the condition
$f \approx 2 \times 10^{-3} (R_jC_j)^{-1}$ 
ensures the level $|\delta I_{\rm{mis}}|/I < 10^{-8}$.
For the R-pump with $R = 50$~k$\Omega$,  this condition reads
$f \approx 2 \times 10^{-3} (R'C_j)^{-1} \approx 30$~MHz
for $R_j = 200$~k$\Omega$ and $C_j = 150$~aF.
This driving frequency corresponds to $I \approx 5$~pA.
Although the larger values of $R$ apparently lead to an appreciable 
reduction of maximum $f$, the use of more low-ohmic junctions
($R_j \sim 50-100$~k$\Omega$) might possibly be
allowed \cite{Comm2}, and this may provide a substantial 
velocity gain in the cycling rate.  

As to the errors associated with thermally activated 
tunnel events with  the magnitude 
$|\delta I_{\rm{th}}|/I \sim \exp (- \tilde{E}/k_B T)$, 
they can also be made sufficiently low. 
With the state-of-the-art junctions of $C_j \approx 150$~aF
and a typical effective electron temperature of the small 
islands of about 60~mK \cite{Est}, we can 
reduce these errors below the level $10^{-8}$. 
Due to an appreciably larger volume of the resistors and their
connection to external leads, their electron temperature
is expected to be even lower. A possible excess shot noise 
in our regime of zero voltage (thermodynamic equilibrium) 
apparently is zero. 

Although the accurate evaluation of cotunneling
relies on simulations based on the theory describing the practical 
case of the $RC$-line, we present here an estimate for the pure
resistance of the moderate value of $R = 50$~k$\Omega$, relying 
on the experimental fact that the effects of this resistance 
and of the $RC$-line on cotunneling are similar.
For the aforementioned set of parameters, 
$viz.$ $C_j = 150$~aF, $R_j = 100$~k$\Omega$, $T = 60$~mK 
and $V = |V_L-V_R| < k_BT/e \approx 5~ \mu$V, formulas Eq.(9)
of paper \cite{GolZai} yield for $N=3$ the rate of cotunneling
through all three junctions,
$\Gamma^{(3)}_{\rm{cot}} \sim 4 \times 10^{-5}$~s$^{-1}$.
This yield 
$|\delta I^{(3)}_{\rm{cot}}|/I = \Gamma^{(3)}_{\rm{cot}}/f  \sim 10^{-11}$
for $f = 30$~MHz. 

When electron is
transferred across one of the junctions the cotunneling across two 
other junctions (in the opposite direction) with the 
rate $\Gamma^{(2)}_{\rm{cot}}$ can occur instead. For this
process the damping effect of the resistors is attenuated by
factor of $\kappa^2 = (2/3)^2 \approx 0.44$ determined by
the ratio of capacitances of the whole chain and of the pair of
junctions. \cite{IngNaz}
Evaluating $\Gamma^{(2)}_{\rm{cot}}$ along the lines of 
Refs. \cite{GolZai} and \cite{Pothier} leads for the
aforementioned parameters to the relative 
errors of pumping $\sim 10^{-8}$.

We conclude that the R-pump with realistic 
parameters combines the potentially 
high accuracy, normally achievable in the many-junction ($N=$ 6-7) 
devices, and the attractive simplicity of the three-junction device.
Its implementation is feasible within the framework of
present-day fabrication technology and the standard
measuring facilities for SET, which assume, in particular, a reliable 
filtering of the microwave frequency noise (see, for 
instance, Ref. \cite{ThCoax}). The work on the Al-Cr R-pump is 
presently in progress at PTB. \cite{Rpump}

\section{Conclusion}

We have shown that compact ($\sim$ several $\mu$m) 
Cr resistors of a magnitude up to 100~k$\Omega$ 
connected in series to a SET device can drastically improve 
the characteristics of this device, considerably suppressing
the cotunneling effect. Although our resistors had a
substantial self-capacitance, the experimental data obtained are 
in a good agreement with theoretical predictions made for the
pure ohmic resistance.
Such a resistor can replace 
$\Delta N = z \equiv R/R_k$ junctions
in a SET circuit in which small cotunneling leakage is crucial.
In particular, the R-SETs have a sharp Coulomb blockade
corner in their $I-V$ curves and, hence, they exhibit
excellent modulation of voltage by gate at a very low level of
bias current $I$.

We showed that the implementation of a three-junction 
electron pump with on-chip resistors is feasible with the present
fabrication technology.  
The accuracy of the controlled transfer of single 
electrons in this R-pump ($|\delta I|/I \leq 10^{-8}$) 
is sufficient for important applications in metrology, 
for example, for the cryogenic capacitor charging 
experiment \cite{Kell}.

The obvious advantage of the R-pump consists in the 
minimum number of islands, namely $N-1=2$. 
Besides the simplicity of the rf-drive (two phase-shifted 
signals only) this design drastically simplifies the adjustment 
of the offset charges on the islands by the gates. 
Moreover, a drift of these charges due to slow recharging
processes in a substrate causing detuning of the pump  
is potentially smaller in the case of two islands of the R-pump.  

\acknowledgments

The authors wish to thank V.~A.~Krupenin for his assistence 
in the measurements. Useful discussions on the initial stage of this
work with K.~Flensberg, A.~A.~Odintsov, F.~Liefrink and 
P.~Teunissen are gratefully 
acknowledged. The work is supported in part by the German 
BMBF (Grant No. 13N7168) and 
the EU (MEL ARI Research Project $22953 -$ CHARGE).

\begin{figure}
\caption{SEM image (a) and the electric measurement 
diagram (b) of R-SET, i.e. the Al transistor 
equipped with 4 identical Cr resistors. This arrangement is
used for 4-point measurements
of the transistor as well as for direct testing of the 
resistors.}
\label{SEM}
\end{figure}

\epsfbox[0 0 30 50]{f2.eps}

\begin{table}
\caption{Parameters of the transistor samples: 
$R_{\Sigma}$ is the sum tunnel resistance;  $C_{\Sigma}$,
the total capacitance of the island.  The 
evaluation of $C_{\Sigma}$ for sample 1 was straightforwardly
made from the offset voltage 
$V_{0}=e/C_{\Sigma}$, while for samples 2 
and 3 the finite environmental impedance corrections
were taken into account. The current unit 
$I_0 = e \, R_k / (\pi ^3 R_{\Sigma}^2 C_{\Sigma})$.
} 
\label{table1}
\begin{tabular}{cccc}
Sample&1&2&3\\ \tableline
Resistance $R_{\Sigma}$ \ [k$\Omega$]&
$240$&$280$&$115$\\ \tableline
Capacitance $C_{\Sigma}$ \ [fF]&
$0.26$&$0.29$&$0.37$\\ \tableline
Voltage unit $V_0$ \ [$\mu$V]&
$610$&$550$&$430$\\ \tableline
Current unit $I_0$ \ [pA]&
$28$&$18$&$8.5$\\ \tableline
Resistance $R$ \ [k$\Omega$]&
$-$&$40$&$80$\\ \tableline
$V_{RC} = \hbar\Omega_{RC}/e$ \ [$\mu$V]&
$-$&$30$&$15$\\
\tableline
$V_c = \hbar\Omega_c/e$ \ [$\mu$V]&
$-$&$70$&$140$\\
\end{tabular}
\end{table}


\begin{references}

\bibitem{AvLik} D.~V.~Averin and K.~K.~Likharev, J.\ Low\
Temp.\ Phys. {\bf 62}, 345 (1986).
\bibitem{Dels} P.~Delsing, K.~K.~Likharev, L.~S.~Kuzmin
and T.~Claeson, Phys.\ Rev.\ Lett.\ {\bf 63}, 1180 (1989).
\bibitem{Clel} A.~N.~Cleland, J.~M.~Schmidt and J.~Clarke, 
Phys.\ Rev.\ Lett.\ {\bf 64}, 1565 (1990).
\bibitem{KuzPash} L.~S.~Kuzmin and Yu.~A.~Pashkin, Physica B 
{\bf 194-196}, 1713 (1994).
\bibitem{Joyez} P.~Joyez, D.~Esteve and M.~H.~Devoret, 
Phys.\ Rev.\ Lett.\ {\bf 80}, 1956 (1998).
\bibitem{Zheng} W.~Zheng, J.~R.~Friedman, D.~V.~Averin, 
S.~Han and J.~E.~Lukens,
Solid\ State\ Commun.\ {\bf 108}, 839 (1998).
\bibitem{Ing} G.~L.~Ingold, P.~Wyrowski and H.~Grabert, 
Z.\ Phys.\ B {\bf 85}, 443 (1991).
\bibitem{LikZor} K.~K.~Likharev and A.~B.~Zorin, J.\ Low\
Temp.\ Phys. {\bf 59}, 347 (1985).
\bibitem{KuzHav} L.~S.~Kuzmin and D.~B.~Haviland, 
Phys.\ Scripta {\bf T42}, 171 (1992).
\bibitem{AvOd} D.~V.~Averin and A.~A.~Odintsov, 
Phys.\ Lett.\ A {\bf 140}, 251 (1989).
\bibitem{OBS} A.~A.~Odintsov, V.~Bubanja and 
G.~Sch\"on, Phys.\ Rev.\ B {\bf 46}, 6875 (1992).
\bibitem{GolZai} D.~S.~Golubev and A.~D.~Zaikin, 
Phys.\ Lett.\ A {\bf 169}, 475 (1992).
\bibitem{NiemDol} J.~ Niemeyer, PTB-Mitt.\ {\bf 84}, 251 (1974); 
G.~J.~Dolan, Appl.\ Phys.\ Lett.\ {\bf 31}, 337 (1977).
\bibitem{IngNaz} G.~L.~Ingold and Yu.~V.~Nazarov, in \it {Single 
Charge Tunneling,} \rm edited by H.~Grabert and M.~H.~Devoret 
(Plenum, New York, 1992), Chapter 2, p.21.
\bibitem{Kuz} L.~S.~Kuzmin, Yu.~V.~Nazarov, D.~B.~Haviland,
P.~Delsing and T.~Claeson, Phys.\ Rev.\ Lett.\ {\bf 67}, 1161 (1991).
\bibitem{Geer} L.~J.~Geerligs, D.~V.~Averin  
and J.~E.~Mooij, Phys.\ Rev.\ Lett. {\bf 65},  3037 (1990).
\bibitem{Trap} S.~V.~Lotkhov, H.~Zangerle, A.~B.~Zorin 
and J.~Niemeyer, Appl.\ Phys.\ Lett.\ {\bf 75}, 2665 (1999).
\bibitem{4trap} T.~A.~Fulton, P.~L.~Gammel, and L.~N.~Dunkleberger,
Phys.\ Rev.\ Lett. {\bf 67},  3148 (1991); P.~Lafarge, P.~Joyez, H.~Pothier, 
A.~Cleland, T.~Holst, D.~Esteve, C.~Urbina and 
M.~H.~Devoret, C.\ R.\ Acad.\ Sci.\ Paris {\bf 314}, 883 (1992).
\bibitem{Kell} M.~W.~Keller, J.~M.~Martinis, N.~M.~Zimmerman
and A.~H.~Steinbach, Appl.\ Phys.\ Lett.\ {\bf 69},  1804 (1996). 
\bibitem{Dress} P.~D.~Dresselhaus, L.~Ji, S.~Han, J.~E.~Lukens
and K.~K.~Likharev, Phys.\ Rev.\ Lett. {\bf 72},  3226 (1994).
\bibitem{Est} D.~Esteve in \it {Single Charge 
Tunneling,} \rm edited by H.~Grabert and M.~H.~Devoret 
(Plenum, New York, 1992), Chapter 3, p.108.
\bibitem{Comm2} As the experiment on the four-junction
R-trap showed \cite{Trap}, rather low tunnel resistances 
of junctions $R_j \sim 70$~k$\Omega$ did not lead to a noticeable 
charge leakage in the chain.
\bibitem{Pothier} H.~Pothier, Ph.D. thesis, Paris VI University, 1991.
\bibitem{ThCoax} A.~B.~Zorin, Rev.\ Sci.\ Instrum.\ {\bf 66}, 4296 (1995).
\bibitem{Rpump} S.~V.~Lotkhov, S.~A.~Bogoslovsky, A.~B.~Zorin 
and J.~Niemeyer, to be presented at Conference on Precision
Electromagnetic Measurements CPEM 2000 (14-19 May 2000,
Sydney, Australia).

\end{references}
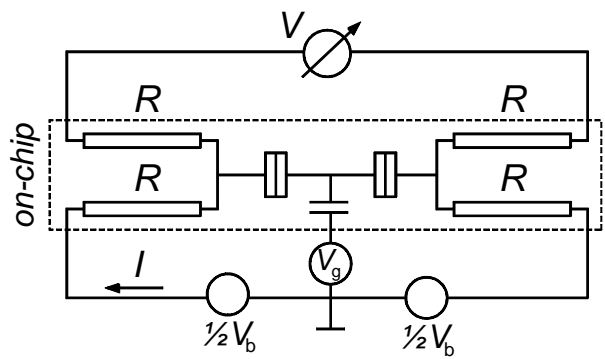
\end{document}